\documentclass[twocolumn]{aastex62}

\received{---}
\revised{---}
\accepted{---}
\submitjournal{ApJ}

\shorttitle{Massive Quenching Galaxy at $z=4.01$}
\shortauthors{Tanaka et al.}

\begin{document}

\title{Stellar Velocity Dispersion of a Massive Quenching Galaxy at $z=4.01$}

\correspondingauthor{Masayuki Tanaka}
\email{masayuki.tanaka@nao.ac.jp}

\author{Masayuki Tanaka}
\affil{National Astronomical Observatory of Japan, 2-21-1 Osawa, Mitaka, Tokyo 181-8588, Japan}
\affil{Department of Astronomical Science, The Graduate University for Advanced Studies, SOKENDAI, Mishima,411-8540 Japan}

\author{Francesco Valentino}
\affil{Cosmic Dawn Center (DAWN)}
\affil{Niels Bohr Institute, University of Copenhagen, Lyngbyvej 2, DK-2100 Copenhagen, Denmark}

\author{Sune Toft}
\affil{Cosmic Dawn Center (DAWN)}
\affil{Niels Bohr Institute, University of Copenhagen, Lyngbyvej 2, DK-2100 Copenhagen, Denmark}

\author{Masato Onodera}
\affil{Subaru Telescope, National Astronomical Observatory of Japan, 650 N Aohoku Pl, Hilo, HI96720}
\affil{Department of Astronomical Science, The Graduate University for Advanced Studies, SOKENDAI, Mishima,411-8540 Japan}

\author{Rhythm Shimakawa}
\affil{Subaru Telescope, National Astronomical Observatory of Japan, 650 N Aohoku Pl, Hilo, HI96720}

\author{Daniel Ceverino}
\affil{Universidad Aut\'onoma de Madrid, Ciudad Universitaria de Cantoblanco, 28049 Madrid}

\author[0000-0002-9382-9832]{Andreas L. Faisst}
\affil{IPAC, California Institute of Technology, 1200 East California Boulevard, Pasadena, CA 91125, USA}

\author{Anna Gallazzi}
\affil{INAF – Osservatorio Astrofisico di Arcetri, Largo Enrico Fermi 5, I-50125 Firenze, Italy}

\author{Carlos G\'omez-Guijarro}
\affil{Cosmic Dawn Center (DAWN)}
\affil{Niels Bohr Institute, University of Copenhagen, Lyngbyvej 2, DK-2100 Copenhagen, Denmark}
\affil{Laboratoire AIM-Paris-Saclay, CEA/DSM-CNRS-Universite Paris Diderot, Irfu/Service d’Astrophysique, CEA Saclay, Orme des Merisiers, F-91191 Gif-sur-Yvette, France}

\author{Mariko Kubo}
\affil{National Astronomical Observatory of Japan, 2-21-1 Osawa, Mitaka, Tokyo 181-8588, Japan}

\author{Georgios E. Magdis}
\affil{Cosmic Dawn Center (DAWN)}
\affil{Niels Bohr Institute, University of Copenhagen, Lyngbyvej 2, DK-2100 Copenhagen, Denmark}
\affil{DTU Space, National Space Institute, Technical University of Denmark, Elektrovej 327, DK-2800 Kgs. Lyngby, Denmark}

\author{Charles L. Steinhardt}
\affil{Cosmic Dawn Center (DAWN)}
\affil{Niels Bohr Institute, University of Copenhagen, Lyngbyvej 2, DK-2100 Copenhagen, Denmark}

\author{Mikkel Stockmann}
\affil{Cosmic Dawn Center (DAWN)}
\affil{Niels Bohr Institute, University of Copenhagen, Lyngbyvej 2, DK-2100 Copenhagen, Denmark}

\author{Kiyoto Yabe}
\affil{Kavli Institute for the Physics and Mathematics of the Universe (Kavli IPMU, WPI), University of Tokyo, Chiba 277-8582, Japan}

\author{Johannes Zabl}
\affil{Univ Lyon, Univ Lyon1, Ens de Lyon, CNRS, Centre de Recherche Astrophysique de Lyon UMR5574, F-69230 Saint-Genis-Laval, France}

\begin{abstract}
  We present the first stellar velocity dispersion measurement of a
  massive quenching galaxy at $z=4.01$.
  The galaxy is first identified as
  a massive $z\geq4$ galaxy with suppressed star formation from photometric redshifts based
  on deep multi-band data.  A follow-up spectroscopic observation with MOSFIRE on
  Keck revealed strong multiple absorption features, which are identified as Balmer
  lines, giving a secure redshift of $z=4.01$.
  This is the most distant quiescent galaxy known to date.
  Thanks to the high S/N of the spectrum,
  we are able to estimate the stellar velocity dispersion, $\sigma=268\pm59\rm\ km\ s^{-1}$,
  making a significant leap from the previous highest redshift measurement
  at $z=2.8$.
  Interestingly, we find that the velocity dispersion is consistent with that of massive galaxies today, implying
  no significant evolution in velocity dispersion over the last 12 Gyr.
  Based on an upper limit on its physical size from deep optical images
  ($r_{\rm eff}<1.3$ kpc), we find that its dynamical mass is consistent with the stellar mass
  inferred from photometry.  Furthermore, the galaxy is located on the mass fundamental plane
  extrapolated from lower redshift galaxies.  The observed no strong evolution in $\sigma$
  suggests that the mass in the core of massive galaxies does not evolve significantly,
  while most of the mass growth occurs in the outskirts of the galaxies, which also increases
  the size.  This picture is consistent with a two-phase formation scenario in which
  mass and size growth is due to accretion in the outskirts of galaxies via mergers.
  Our results imply that the first phase may be completed as early as $z\sim4$.
\end{abstract}


\section{Introduction}
\label{sec:intro}

The majority of massive galaxies in the local Universe have not been actively forming stars
for $\gtrsim10$ billion years.  Their spectroscopic properties are consistent with
an intense starburst occurred in the early Universe followed by passive evolution
(e.g., \citealt{thomas05,gallazzi05,renzini06}).
The recent advent of sensitive near-IR spectrographs has allowed us to reach close to
their primary formation epoch. There are a handful of massive quiescent galaxies confirmed
out to $z=3.7$ \citep{glazebrook17,schreiber18}.  Detailed photometric analyses suggest that they
indeed seem to form in a short and intense starburst, followed by rapid quenching
\citep{schreiber18}.  However, the physics of this entire process still remains unclear.

A key parameter for characterizing quiescent galaxies is the stellar velocity dispersion,
which is an integrated motion of stars along the line of sight.  It exhibits tight
correlations with other fundamental properties of galaxies (e.g., the fundamental plane; \citealt{djorgovski87}).
It is also the best predictor of stellar age with no residual correlation between size
and age at fixed velocity dispersion \citep{vanderwel09,graves09}.
The velocity dispersion measurement is observationally challenging due to the demanding
S/N ratio of rest-frame optical spectra.  It has been measured out to $z=2.8$ with
the help of gravitational lensing effects \citep{hill16}.  Although quiescent galaxies have been
confirmed at higher redshifts, their dynamical properties remain unknown.  This is unfortunate
because dynamical information may also hold a key to understanding how these galaxies form.

This paper presents the  spectroscopic confirmation of a massive quenching galaxy at
$z=4.01$, the most distant galaxy with suppressed star formation rate (SFR) known to date.
The paper further presents the measurement of its stellar velocity dispersion,
opening a new window to explore dynamical properties of massive galaxies at $z\sim4$.
A companion paper \citep{valentino19}
discusses star formation histories and progenitors of this and other galaxies at $z\sim4$ in detail.
We first
summarize our observation in Section 2, and then present spectral analyses in Section 3.
Measurements of the physical size and discussions of its dynamical properties are given in
Sections 4 and 5, respectively.  Finally, we discuss implications of our results and
conclude the paper in Section 6.
We adopt $\rm H_0=70\ km\ s^{-1}\ Mpc^{-1}$, $\rm \Omega_M=0.3$,
and $\rm \Omega_\Lambda=0.7$, unless otherwise stated.  Magnitudes are in the AB system.

\section{Observation}
\label{sec:observation}

\subsection{Target Selection}
\label{sec:target_selection}

We select massive quiescent galaxy candidates located at $z\sim4$ using deep multi-wavelength data available
in Subaru/XMM-Newton Deep Field.  We have compiled $uBVRizJHK[3.6][4.5][5.8][8.0]$ photometry measured
in a consistent manner and applied a photometric redshift code from \citet{tanaka15} to infer redshifts
as well as SFR and stellar mass.  Further details are given in \citet{kubo18}.
We define galaxies with $1\sigma$ upper limit of specific SFR below $10^{-9.5}\rm\ yr^{-1}$ as quiescent galaxies.
These candidates are typically $K\sim24$.
Among them, there is one outstandingly bright
galaxy with $K=21.9$, which is the subject of the paper.  There is no nearby galaxy or galaxy cluster,
and this object is not strongly lensed.
Fig.~\ref{fig:spectrum} (right panel) shows the SED of the object.
It is a massive galaxy at $z_{phot}\sim4$ exhibiting a prominent Balmer break, which is
indicative of a recent starburst a few hundred Myr ago.  The SFR inferred from the fit
is low for its stellar mass; SFR$=24.0^{+21.7}_{-22.7}$ and $M_*=1.15^{+0.11}_{-0.10}\times10^{11}\rm\ M_\odot$,
giving a specific SFR of $\sim10^{-10}\rm\ yr^{-1}$. The object is about 1~dex below
the sequence of star forming galaxies on a SFR vs. $M_*$ diagram and
it is likely that this galaxy recently quenched (see \citealt{valentino19} for further discussions).

\subsection{Observation and Data Reduction}
\label{sec:data_reduction}

We were allocated the first half nights of the 20--21 December 2018 to follow-up the target
with Keck/MOSFIRE.  The observing conditions were
good and the seeing was around 0.7 arcsec.  Each exposure was 180 seconds long and
the classical ABBA nodding was applied along the slit.  The total integration time was 7.75 hours.
A bright star was put in the mask, which allowed us to keep a track of changes in the observing conditions.
7\% of the exposures with low fluxes from the bright star were excluded due to poor seeing and/or poor alignment.
The data were processed in a standard manner using the MOSFIRE DRP 2018 release. 
A0V stars were observed as spectrophotometric standards twice per night.  The flux calibration vectors
derived from those stars were consistent at a few percent level on both nights. 
The vectors were averaged and the mean flux calibration vector was applied to the extracted 1d spectra.

Fig.~\ref{fig:spectrum} shows the spectrum.  There are four prominent absorption lines, which
were identified as H$\gamma$ to H$\zeta$.  There is also H$\eta$ on the edge.  We confirm the redshift
to be $z=4.012\pm0.001$.
This proves the accuracy of our photometric redshift estimate ($z_{phot}=4.12$).
There is no emission line in the spectrum, and
$1\sigma$ upper limit on (unobscured) SFR from H$\gamma$ is $\lesssim8\rm\ M\odot\ yr^{-1}$.
There is no IR detection either.
All this confirms the quiescent nature of the galaxy (see \citealt{valentino19}).

\section{Spectral Fit}
\label{sec:spectral_fit}

\begin{figure*}
\centering
\includegraphics[width=180mm]{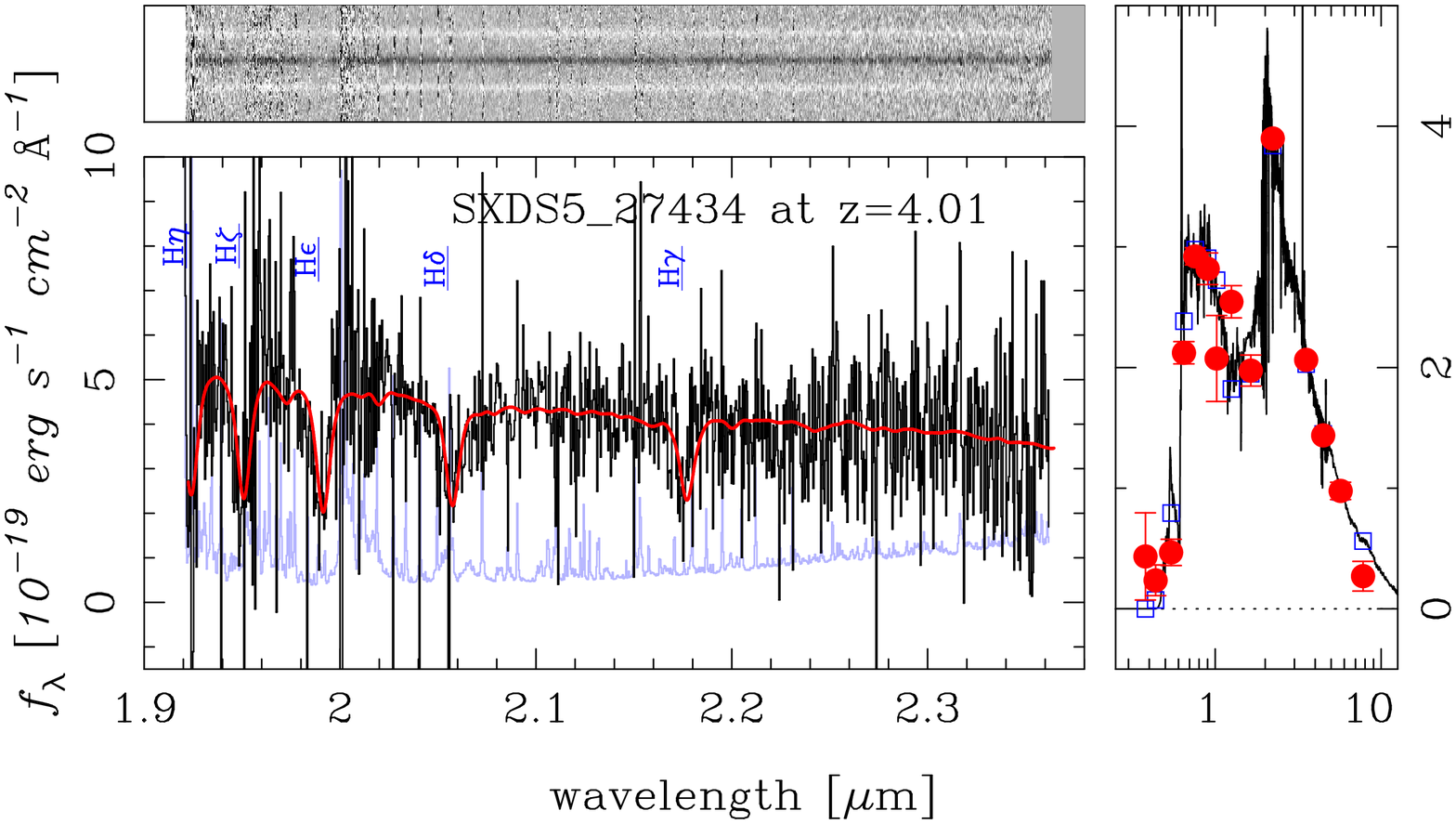}
\caption{
  {\bf Left:}
  MOSFIRE $K$-band spectrum.  The top panel shows the 2D spectrum.  The A and B nods are in white and
  the combined spectrum is in black.  The bottom panel shows the extracted 1D spectrum.
  The black and blue spectra are the object and noise spectra binned over approximately the resolution element, respectively.
  The red spectrum is the best-fit model spectrum from {\tt ppxf}.
  {\bf Right:}
  Broad-band SED of the object.  The red circles and blue squares are the observed and model
  photometry, respectively.  The spectrum is the best-fitting model spectrum from the photometric redshift code
  \citep{tanaka15}.
}
\label{fig:spectrum}
\end{figure*}

We fit the observed MOSFIRE spectrum using {\tt ppxf} \citep{cappellari04,cappellari17}.
We use the simple stellar population models from \citet{vazdekis10} for the fit.  We exclude all models that are older
than the age of the Universe at the redshift of the object.  We also exclude models
with sub-solar metallicity as we are focusing on a massive galaxy.  Since the rest-frame
spectral resolution of the MOSFIRE spectrum is slightly higher than that of the library, we apply
Gaussian smoothing to match the resolution.  We use an additive correction function of
order 1 (i.e., linear) and no multiplicative correction.
Our results are not sensitive to the choices here.

The best-fit spectrum is shown in red in Fig.~\ref{fig:spectrum}.  All observed Balmer absorption
lines are fit very well, and the overall fit has $\chi^2_\nu=1.2$.
In addition, thanks to the high S/N of
the spectrum    ($S/N\sim5$ per resolution element),
we measure a velocity dispersion of $268\pm59\rm\ km\ s^{-1}$.  The uncertainty
here is based on a Monte-Carlo simulation; we perturb the observed spectrum using the noise
spectrum and repeat the fits.  The quoted uncertainty is the 68th percentile of the distribution
from the Monte-Carlo fits.  If we use the Indo-US stellar spectral library \citep{valdes04} instead of Vazdekis
so that we do not need to smooth the MOSFIRE spectrum, we obtain a consistent stellar
velocity dispersion ($\sigma=252\pm77\rm\ km\ s^{-1}$).   We also have confirmed that emission line in-filling
does not affect our measurement either; we repeat the fits by excluding H$\gamma$ and H$\delta$ and obtain
a fully consistent measurement, $\sigma=277\pm58\rm\ km\ s^{-1}$.

\section{Physical Size}
\label{sec:physical_size}

\begin{figure*}
\centering
\includegraphics[width=80mm]{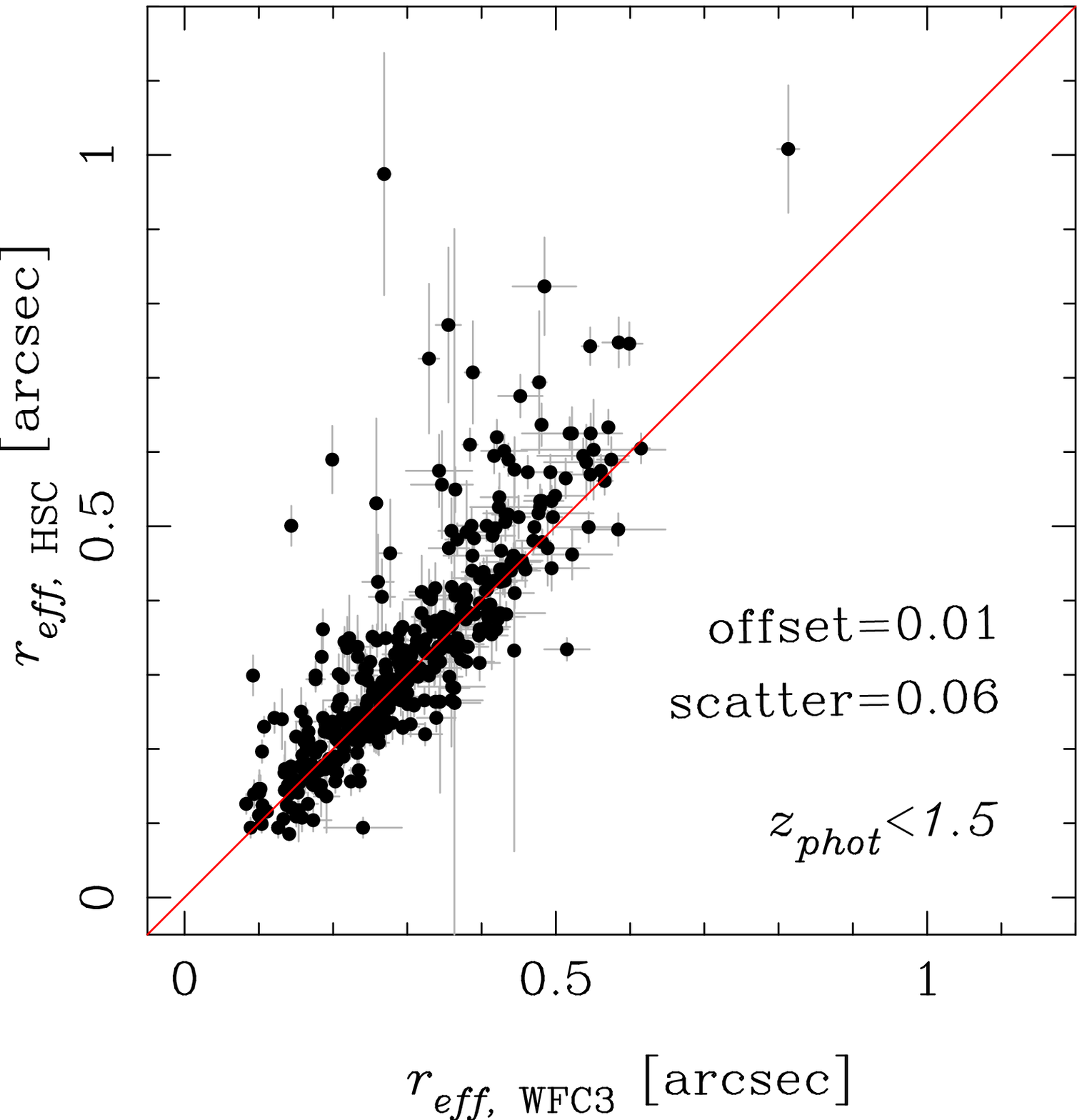}\hspace{0.5cm}
\includegraphics[width=80mm]{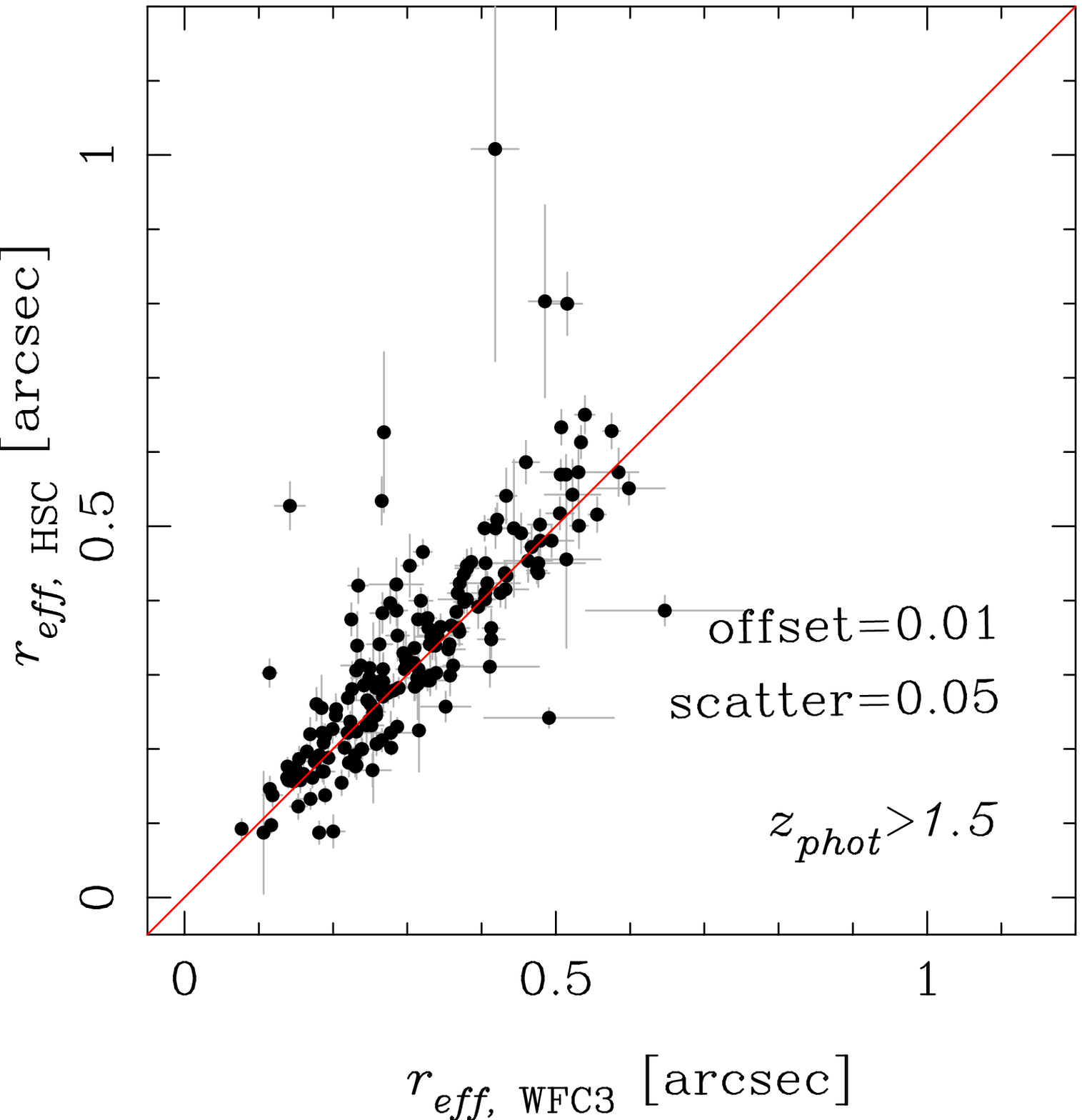}
\caption{
  Size from HSC $i$-band plotted against size from {\it HST} $F160W$.  The plotted are objects with $i$-band magnitudes
  similar to that of the $z=4$ object.  The solid line shows $r_{\rm eff,HSC} = r_{\rm eff,HST}$.
  The left and right panels show objects at $z<1.5$ and $z>1.5$, respectively.  The systematic offset and scatter
  around it are shown in the figures.
}
\label{fig:size_meas}
\end{figure*}

In order to fully utilize the stellar velocity dispersion, we are interested in structural properties
of the $z=4$ galaxy, in particular its size.  The typical rest-frame optical size of $z\sim4$ galaxies is very small
($r_{\rm eff}\sim0.5\rm\ kpc$; \citealt{kubo18}), and adaptive optics assisted observations or high spatial resolution
imaging with the Hubble Space Telescope ({\it HST}) are an ideal way to measure the size of the $z=4$ galaxy.
Unfortunately, there is no existing {\it HST} data for the galaxy and \citet{kubo18} did not observe the galaxy with AO.

However, a useful upper limit on the size can still be obtained from the ground-based data.
We use deep optical data from Hyper Suprime-Cam Subaru Strategic Program (HSC-SSP; \citealt{aihara18,aihara19})
and demonstrate how well we can reproduce the sizes measured from {\it HST}.
The $i$-band image from HSC-SSP processed with the pipeline designed for LSST \citep{juric17,bosch18,ivezic19}
is used here to measure sizes due to superior seeing than the other bands.
We retrieve objects with similar $i$-band magnitudes to the $z=4$ galaxy ($i\sim24.3$) in UltraDeep-SXDS, which is the same field
as UDS, in which the $z=4$ galaxy is located, and run {\tt galfit} \citep{peng02,peng10} to measure their effective radii
adopting the Sersic profile.
We use the variance image to generate a sigma map and use the 'coadd PSF' \citep{bosch18} as an input PSF image.
The PSF has FWHM$=0.65$ arcsec.  We perform a Monte-Carlo run by slightly perturbing the initial centroid,
position angle, effective radius and brightness.  We fix the Sersic index to a randomly drawn value between 0.5 and 4
in each run.
We use the fit with the smallest $\chi^2$ as the best estimate and $\Delta\chi^2<1$ as the 68th percentile interval.

Fig.~\ref{fig:size_meas} compares the size measurements between HSC and {\it HST}.  The {\it HST} sizes are taken from \citet{vanderwel14}
and are measured in the WFC3/F160W filter.
We split the sample into low and high redshift ranges using the photo-$z$ described in Section \ref{sec:target_selection}.
These two panels give a quantitative estimate of the rest-frame wavelength dependence of the size measurements.
Interestingly, we observe equally good correlation between the two measurements down to $\sim0.1$ arcsec in both plots.
If we define outliers as $|r_{\rm eff,HSC}-r_{\rm eff,HST}|>0.2$ arcsec, the outlier rate is about 4\% in both plots.
This good correlation is likely due to the depth of the HSC data (the $5\sigma$ depth is $\sim27$~mag) and also to the good PSF model.
Other studies also have shown that sizes can be measured to $\sim0.2$ FWHM \citep{vanderwel14}
for high S/N and adequate spatial sampling.
It is encouraging that there seems no major evidence for large morphological k-corrections in the size measurements.
We note that \citet{vanderwel14} also find that the wavelength dependence of size is not very strong for quiescent galaxies.

\begin{figure*}
\centering
\includegraphics[width=50mm]{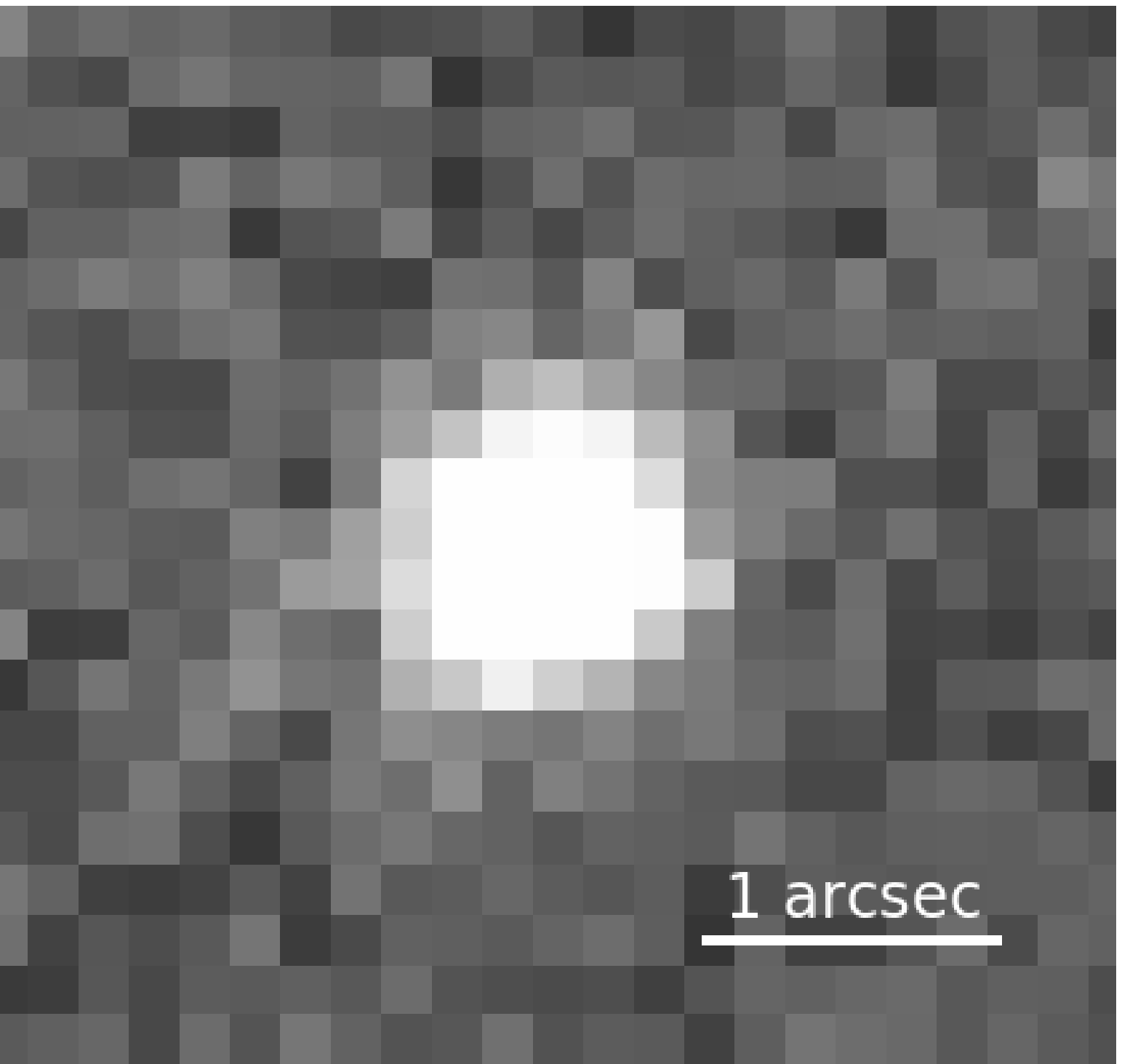}
\includegraphics[width=50mm]{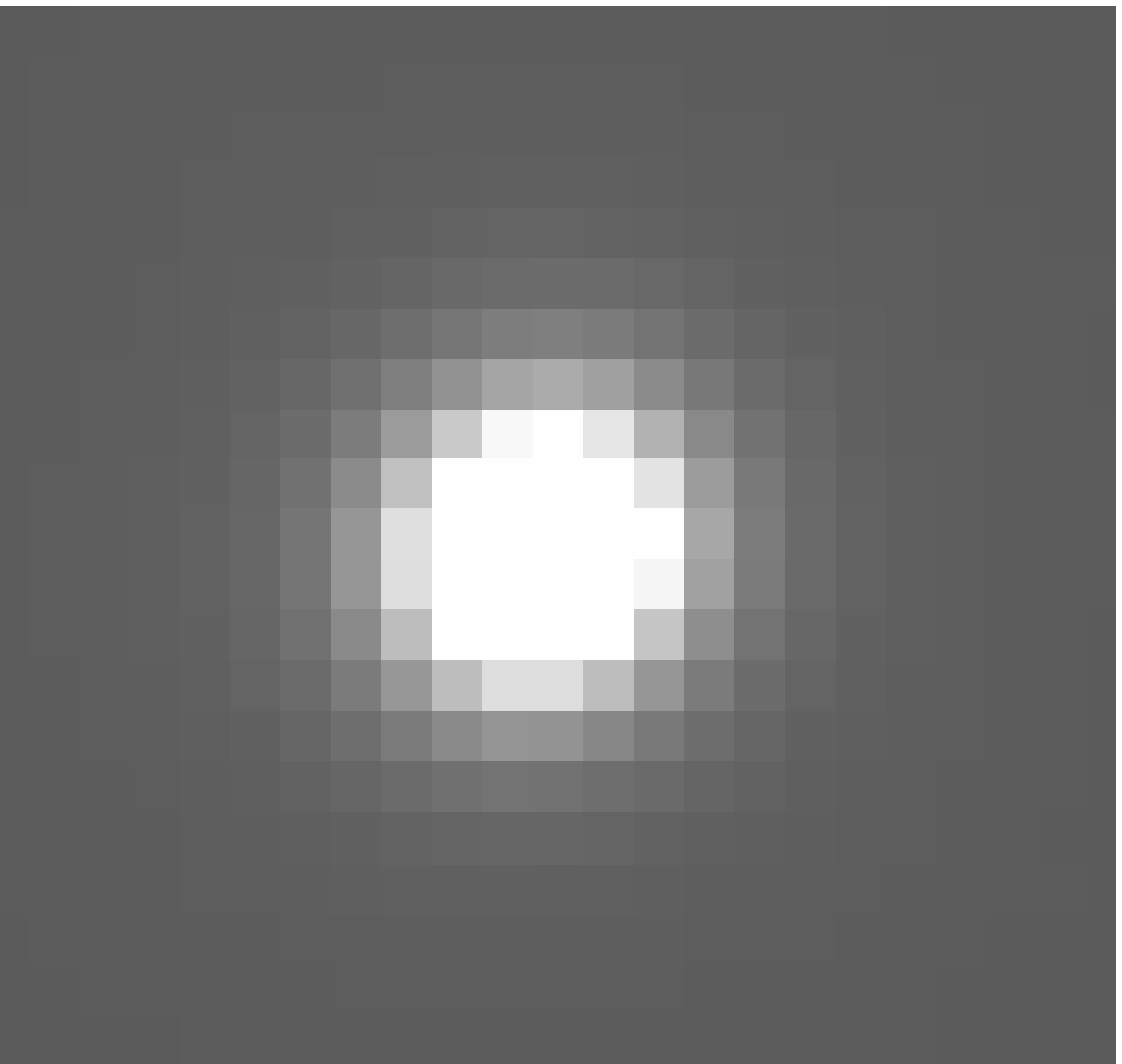}
\includegraphics[width=50mm]{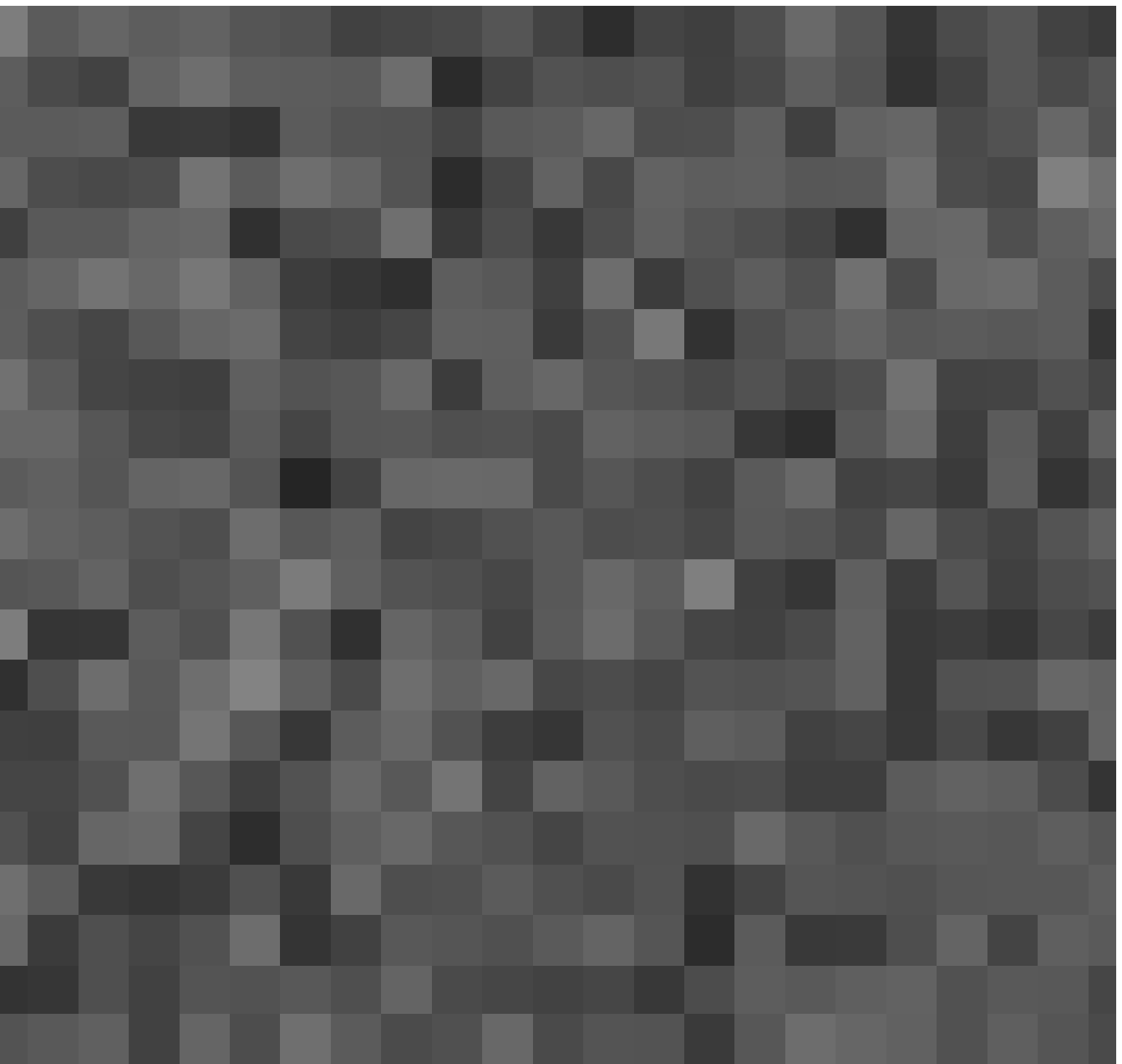}
\caption{
  The panels show, the $i$-band image from HSC, the best-fit {\tt galfit} model, and residual, from left to right.
  The flux scales are the same in all the panels.
}
\label{fig:galfit}
\end{figure*}

Adopting this size measurement procedure for the $z=4$ galaxy, we find the best size estimate
of $0.11 \pm 0.03$ arcsec ($0.76\pm0.20$ kpc).  This size is fully consistent with the typical size of $z\sim4$
massive quiescent galaxies from \citet{kubo18}.
Fig.~\ref{fig:galfit} shows the object, best-fit model, and residual images, demonstrating an excellent fit.
Figs.~\ref{fig:size_meas} and \ref{fig:galfit} suggest that our estimate here is likely reasonable, but
to be fully conservative, we primarily use the upper limit on the size in what follows; $0.76+0.20$ (random) $+0.35$
(systematic corresponding to 0.05 arcsec observed scatter) $=1.3\rm\ kpc$.

\begin{figure*}
\centering
\includegraphics[width=80mm]{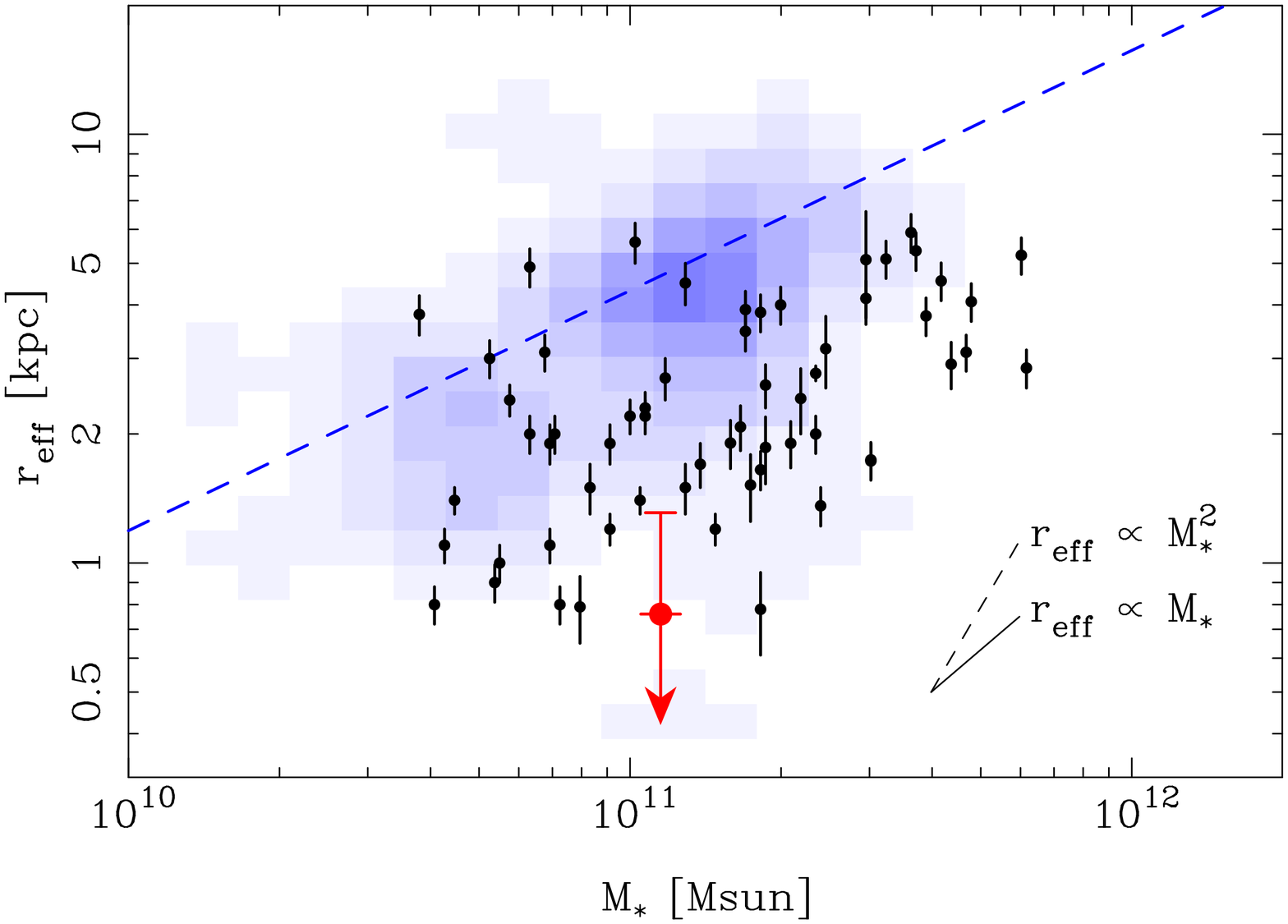}\hspace{0.5cm}
\includegraphics[width=83mm]{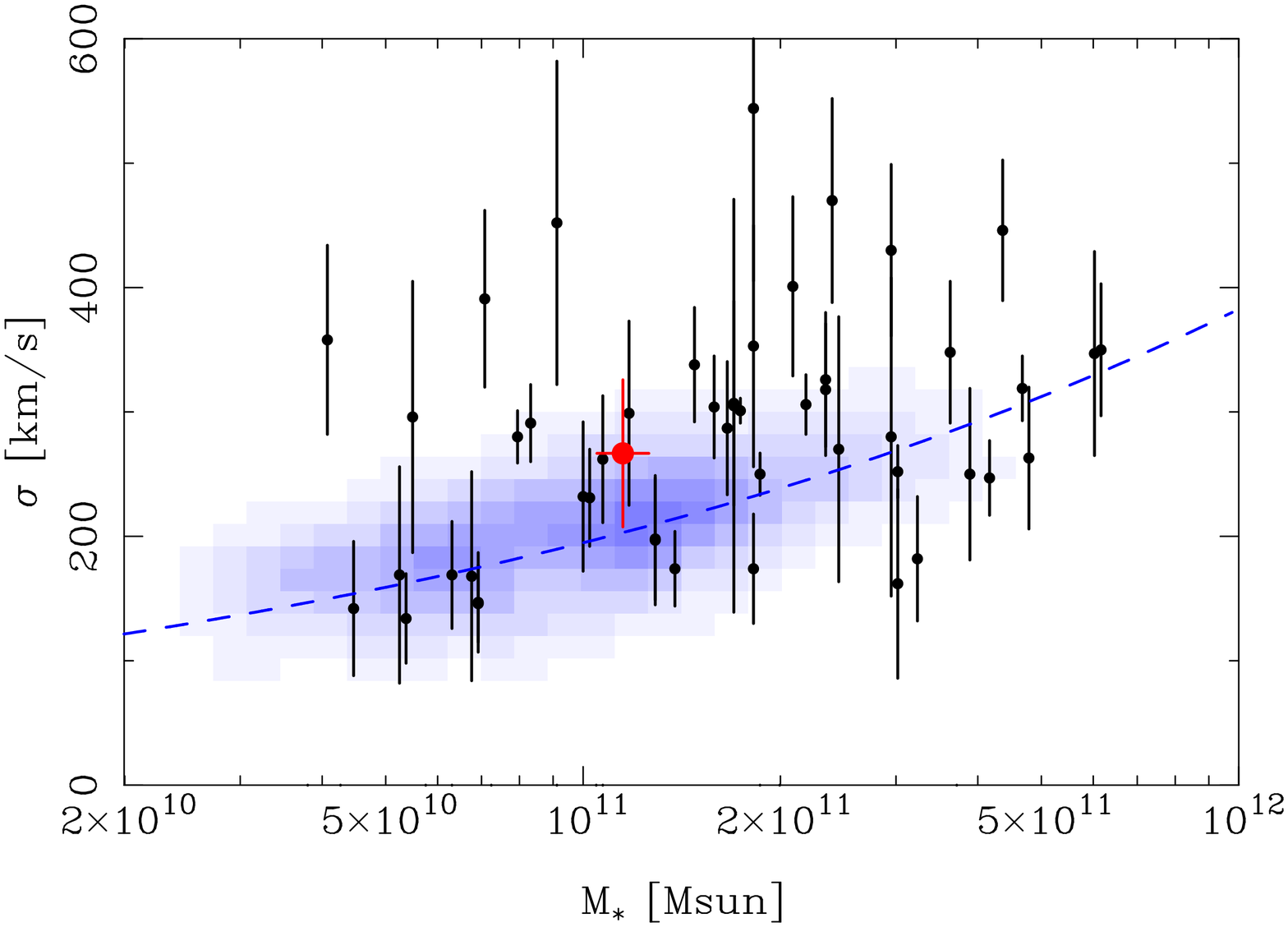}
\caption{
  {\bf Left:} Size plotted against stellar mass. 
  The $z=4$ galaxy is shown in red. The black points are quiescent galaxies at $z>1.5$
  drawn from the literature; compilation by \citet{vandesande13}, which includes
  \citet{vandokkum09,onodera12,toft12,bezanson13a}, and newer data from \citet{belli17} and Stockmann et al. (submitted).
  Their mean redshift is around 2.
  The shades show data from the LEGA-C survey, which covers $0.6\lesssim z\lesssim1.0$.
  The dashed line shows the local relation from \citet{shen03}
  based on data from the Sloan Digital Sky Survey (SDSS; \citealt{york00,strauss02}).
  The solid and dotted lines in the bottom-right corner shows evolutionary tracks with $r_{\rm eff}\propto M_*$ and
  $r_{\rm eff}\propto M_*^2$, respectively.  They represent the major and minor merger tracks.
  {\bf Right:} Stellar velocity dispersion plotted against stellar mass.  The meaning of the symbols are the same as
  in the left figure.
  The blue dashed curve is a fit to SDSS galaxies \citep{zahid16}.
}
\label{fig:re_vs_mstar}
\end{figure*}

The left panel of Fig.~\ref{fig:re_vs_mstar} shows $r_{\rm eff}$ against stellar mass.
In addition to the $z=4$ galaxy, the figure also includes lower redshift galaxies for
comparison; quiescent galaxies at $z\sim2$ from a compilation of literature and
$0.6\lesssim z \lesssim1.0$ galaxies from the LEGA-C survey \citep{vanderwel16,straatman18} cross-matched with
the size measurements from \citet{vanderwel14}.  Quiescent galaxies among the LEGA-C galaxies are selected
using the multi-band classification from \citet{laigle16}.
The $z=4.01$ galaxy is compact for its stellar mass, and it is physically smaller than $z\sim2$
and other lower redshift galaxies. This clear redshift trend is consistent with \citet{kubo18}.

In contrast, the stellar velocity dispersion of the $z=4$ galaxy plotted
in the right panel is largely consistent with those of the lower redshift galaxies.
There is a possible hint that low redshift galaxies from LEGA-C and SDSS
have a slightly lower velocity dispersion.
However, the difference, if any, is fairly modest.
This indicates that the stellar velocity dispersion has not significantly evolved
over the last 12 Gyr.  This is an intriguing result because the size and mass
are known to evolve significantly over this time period.

\section{Dynamical Analysis}
\label{sec:dynamical_analysis}

Using the size from
the previous section, we estimate the galaxy's dynamical mass as

\begin{equation}
  M_{dyn} = \frac{\beta(n)\ \sigma^2\ r_{\rm eff}}{G},
\end{equation}

\noindent
where $\beta$ is $\beta(n)=8.87-0.831n + 0.024n^2$ \citep{cappellari06}.  As discussed in the same paper,
a constant $\beta=5.0$ also works well.  Given the unconstrained $n$ for our object, we adopt $\beta=5$ here,
which yields $M_{dyn}=6.3\times10^{10}\rm\ M_\odot$ with an upper limit of $1.2\times10^{11}\ \rm M_\odot$.

\begin{figure*}
\centering
\includegraphics[width=80mm]{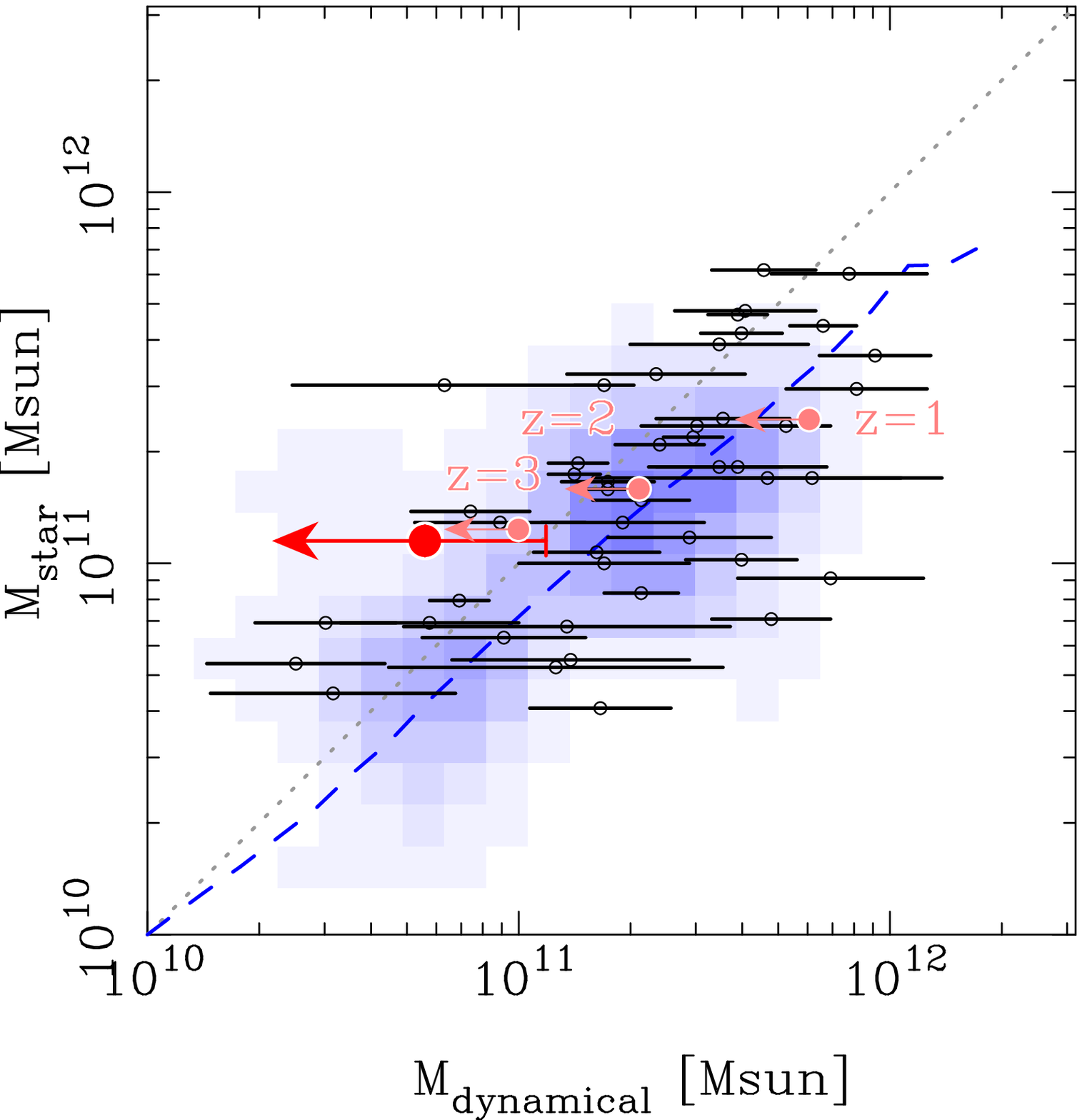}\hspace{0.5cm}
\includegraphics[width=83mm]{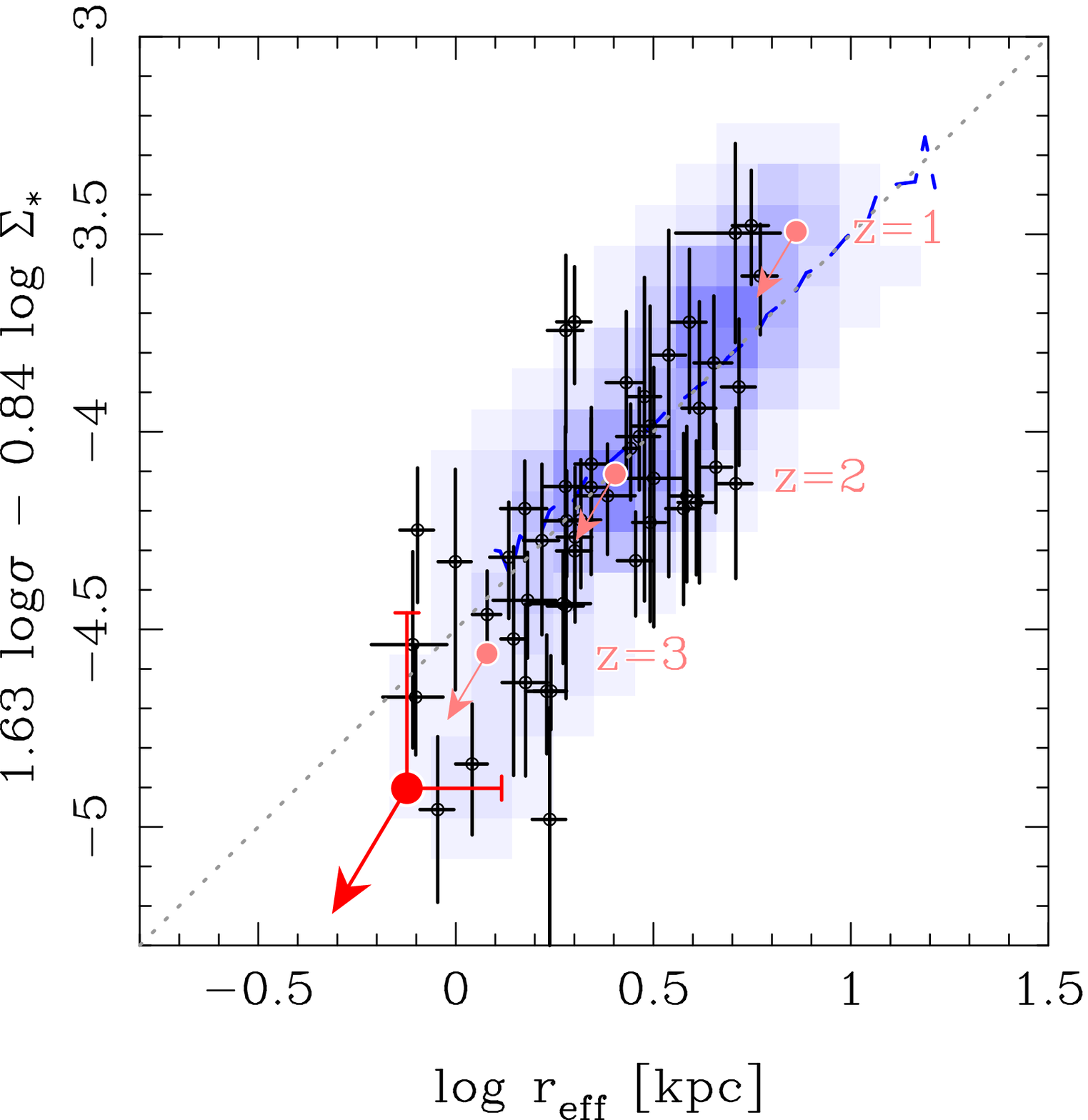}
\caption{
  {\bf Left:} Stellar mass plotted against dynamical mass.  The dotted line shows $M_{*}=M_{\rm dyn}$.
  The black points are the lower redshift objects ($z\sim2$) as in Fig.~\ref{fig:re_vs_mstar}.
  The shades are quiescent galaxies from LEGA-C, and the dashed curve is the running median of the distribution
  of quiescent galaxies with sSFR$<10^{-10}\rm\ M_\odot\ yr^{-1}$ drawn from SDSS DR15 \citep{aguado19}.
  SFR and stellar mass are from Granada FSPS fits \citep{ahn14}.
  The red point is the $z=4$ object.
  The pink points are the expected evolutionary track assuming constant velocity dispersion (see text for details).
  {\bf Right:} 
  Mass fundamental plane.  The dotted line is a fit from \citet{bezanson13b}.
  The other symbols are the same as the left panel.
}
\label{fig:mstar_vs_mdyn}
\end{figure*}

Fig.~\ref{fig:mstar_vs_mdyn} (left panel) compares the stellar mass with the dynamical mass estimated here.
The stellar mass of the $z=4$ object is consistent with the dynamical mass.
We use the Chabrier IMF here \citep{chabrier03}.
This consistency is encouraging because this confirms the accuracy of our photometry-based estimate of stellar mass,
which then implies that assumptions employed such as the initial mass function are reasonable.
The $z=4$ galaxy is expected to increase its size and stellar mass with time through mergers.
\citet{marchesini14} estimated the likely stellar mass growth from abundance matching, and
\citet{kubo18} followed the size growth of these galaxies on the stellar mass evolutionary track from \citet{marchesini14}.
From these results, we can make a prediction for
how the galaxy will evolve in the $M_*$ vs. $M_{dyn}$ plane.  If we assume that the velocity dispersion does not
change with time, which is the assumption also taken by, e.g., \citet{belli17}, the expected location of
the descendant of the $z=4$ galaxy is shown as the pink points (Fig,~\ref{fig:mstar_vs_mdyn}).
The expected evolutionary track shows an increase in the dynamical mass with a smaller increase
in the  stellar mass.  If minor mergers entirely drive the evolution and velocity dispersion does not change
over time, we expect $M_*\propto r_{\rm eff}^{0.5}\propto M_{dyn}^{0.5}$.  As the size growth observed by \citet{kubo18}
is close to the maximum growth rate expected from minor mergers, the evolutionary track is indeed close to
$M_*\propto M_{dyn}^{0.5}$.

In the local Universe, early-type galaxies are known to show
a tight relationship between effective radius, mean intensity, and velocity dispersion (i.e., fundamental plane; \citealt{djorgovski87}).
The evolution of the fundamental plane is mostly due to the evolving stellar mass to luminosity ratio due to
stellar aging.  The mass fundamental plane, which replaces luminosity with stellar mass, has been suggested
in the literature to largely remove that effect \citep{bezanson13b}.
Fig.~\ref{fig:mstar_vs_mdyn} (right panel) shows the mass fundamental plane.

The $z=4$ galaxy is located at the bottom left part of the distribution of the lower redshift galaxies
as expected from its small size and high stellar density. 
The evolutionary track with constant velocity dispersion is along the locations of the lower redshift objects and the galaxy
will likely be among the most massive galaxies at each epoch.
This in turn implies that the velocity dispersion does not evolve significantly.  If it had changed by
a factor of, e.g., 2 by $z=1$, the expected location of the descendant galaxy would be inconsistent
with the observed $z\sim1$ galaxies.

We also briefly discuss the classical fundamental plane. The interpretation of
its evolution is not straightforward, but assuming that the evolution in the fundamental plane is
entirely due to the luminosity evolution for simplicity, we find that the $z=4$ galaxy is $1.70^{+\infty}_{-0.46}$~dex
off in $\log I_{\rm eff}$ in the rest-frame $g$-band from the local relation (recall that we have only
an upper limit on size).
A simple stellar population model from \citet{bruzual03} predict a luminosity evolution of a factor
of $\sim~80$, or equivalently $\sim~1.9$~dex, adopting the luminosity-weighted age of $\sim0.2$ Gyr \citep{valentino19}
for the galaxy.  The fundamental plane evolution therefore is consistent with the pure luminosity evolution,
although the uncertainty here is rather large and we do not discuss further at this point.

\section{Summary and Discussion}
\label{sec:discussion}

We have presented spectroscopic confirmation of a massive galaxy being quenched at $z=4.01$,
which is the most distant example known to date.  Thanks to the high S/N of the spectrum, we are able
to measure its stellar velocity dispersion, $268\pm59\rm\ km\ s^{-1}$.  The size estimate based
on the deep optical data is $0.76\pm0.20\rm\ kpc$ with an upper
limit of $<1.3\rm\ kpc$, which is consistent with the typical size of massive quiescent galaxies
from \citet{kubo18}.  Combining the velocity dispersion and size, we find that the dynamical
mass is consistent with the stellar mass inferred from photometry.  Also, the galaxy is on
the mass fundamental plane and the expected evolutionary path of the galaxy is consistent with
the massive quiescent galaxies at lower redshifts.

The most striking finding of this work is that the stellar velocity dispersion of the massive
galaxy at $z=4.01$ is consistent with that of massive galaxies at lower redshifts.
This is in contrast to the very large velocity dispersion of a $z=2.2$ galaxy reported by \citet{vandokkum09},
$510^{+165}_{-95}\rm\ km\ s^{-1}$.
The $z=4$ galaxy is expected to increase its mass by a factor of $2.1_{-1.0}^{+2.2}$ \citep{marchesini14} and size
by a factor of $10.8_{-3.6}^{+5.3}$ \citep{kubo18} by $z=1$.  However, the stellar velocity dispersion evolution is
significantly weaker; Fig.~\ref{fig:re_vs_mstar} (right) shows that even the most massive galaxies at lower
redshifts have only slightly larger velocity dispersion.  This finding indicates that
the size and mass evolution does not significantly increase the total mass contained in
the core of these massive galaxies.  This has significant implications for how galaxies increase their
size and mass; galaxies do not increase mass equally at all radii, and instead most of
the mass growth occurs in the outer parts perhaps through minor mergers, which also increases the effective radius.
This is fully consistent with the two-phase formation scenario \citep{naab07,oser10}.
In fact, simulations show only a mild evolution in velocity dispersion \citep{oser12},
in agreement with our observation here.
An interesting implication here is that the first of the two phases, namely the formation
of the dense core, may be completed as early as $z\sim4$.

It is, however, not clear whether the $z=4$ galaxy is a dispersion-dominated system.
Recent work finds that massive galaxies at $z\sim2$ exhibit significant rotational motion
\citep{toft17,newman18b}.  Simulations predict a relatively small $V/\sigma$ if massive
galaxies at high redshifts form in starbursts triggered by dissipative mergers \citep{wuyts10}.
The reported large rotational motion favors formation through disk instabilities.
It is possible that the $z=4$ galaxy also rotates, potentially complicating the interpretation above.
But, given the absence of sufficient spatial resolution to resolve any rotational motion
from the MOSFIRE data nor even an ellipticity measurement from the imaging, we have to
await further observations to reveal $V/\sigma$ of the $z=4$ galaxy. 
The large rotational motion, if present, has to be damped at lower redshifts because
the majority of the most massive galaxies in the local Universe are slow rotators (e.g., \citealt{veale17}).
Mergers may be able to do the job \citep{lagos18}.

Prior to this work, the most distant velocity dispersion measurements were made
at $z=2.6-2.8$ with a help of a gravitational lensing effect \citep{hill16,newman18a}.
This work makes a major leap in redshift without lensing and demonstrates that the current facilities
have an ability to confirm redshifts and measure the stellar velocity dispersion of
the brightest galaxies at $z>4$.  Massive galaxies at such redshifts are rare, but
ongoing/upcoming massive imaging surveys will be able to construct a significantly
larger sample, which can then be followed up spectroscopically to further extend the work presented here.

\acknowledgments

This work is supported by JSPS KAKENHI Grant Numbers JP23740144 and JP15K17617.
FV and GEM acknowledges the Villum Fonden research grant 13160 “Gas to stars, stars to dust: tracing star formation across cosmic time”.
FV acnknowledges the Carlsberg Fonden research grant CF18-0388 “Galaxies: Rise And Death”.
The Cosmic Dawn Center (DAWN) is funded by the Danish National Research Foundation under grant No. 140.
ST and GEM acknowledge support from the European Research Council (ERC) Consolidator Grant funding scheme (project ConTExt, grant number: 648179).
MO acknowledges support by KAKENHI JP17K14257.  KY acknowledges support by JSPS KAKENHI Grant Number JP18K13578.
We thank the anonymous referee for a useful report, which helped improve the paper.

The HSC collaboration includes the astronomical communities of Japan and Taiwan, and Princeton University. The HSC instrumentation and software were developed by NAOJ, Kavli IPMU, the University of Tokyo, KEK, ASIAA, and Princeton University. Funding was contributed by the FIRST program from Japanese Cabinet Office, the Ministry of Education, Culture, Sports, Science and Technology, the Japan Society for the Promotion of Science, Japan Science and Technology Agency, the Toray Science Foundation, NAOJ, Kavli IPMU, KEK, ASIAA, and Princeton University.  This paper makes use of software developed for LSST. We thank the LSST Project for making their code available as free software at http://dm.lsst.org. This paper is based in part on data collected at the Subaru Telescope and retrieved from the HSC data archive system, which is operated by Subaru Telescope and ADC at NAOJ. Data analysis was in part carried out with the cooperation of CfCA, NAOJ.

Funding for the SDSS and SDSS-II has been provided by the Alfred P. Sloan Foundation, the Participating Institutions, the National Science Foundation, the U.S. Department of Energy, the National Aeronautics and Space Administration, the Japanese Monbukagakusho, the Max Planck Society, and the Higher Education Funding Council for England. The SDSS Web Site is http://www.sdss.org/.

The SDSS is managed by the Astrophysical Research Consortium for the Participating Institutions. The Participating Institutions are the American Museum of Natural History, Astrophysical Institute Potsdam, University of Basel, University of Cambridge, Case Western Reserve University, University of Chicago, Drexel University, Fermilab, the Institute for Advanced Study, the Japan Participation Group, Johns Hopkins University, the Joint Institute for Nuclear Astrophysics, the Kavli Institute for Particle Astrophysics and Cosmology, the Korean Scientist Group, the Chinese Academy of Sciences (LAMOST), Los Alamos National Laboratory, the Max-Planck-Institute for Astronomy (MPIA), the Max-Planck-Institute for Astrophysics (MPA), New Mexico State University, Ohio State University, University of Pittsburgh, University of Portsmouth, Princeton University, the United States Naval Observatory, and the University of Washington.

\facilities{Keck(MOSFIRE)}

\end{document}